\documentclass[12pt]{article}
\title{Statistical mechanics and thermodynamic limit of self-gravitating fermions in $D$ dimensions}
\author{Pierre-Henri Chavanis}
\date{}

\usepackage{amsthm,amsmath,amssymb,a4wide,psfig}
\def\mb#1{\setbox0=\hbox{$#1$}\kern-.025em\copy0\kern-\wd0
\kern-0.05em\copy0\kern-\wd0\kern-.025em\raise.0233em\box0}

\begin{document}
\maketitle
\vspace*{-1cm}
\begin{center}
 Laboratoire de Physique Th\'eorique,
Universit\'e Paul Sabatier,\\
118 route de Narbonne 31062 Toulouse, France.\\

\vspace{0.5cm}
\end{center}

\begin{abstract}

We discuss the statistical mechanics of a system of self-gravitating
fermions in a space of dimension $D$. We plot the caloric curves of
the self-gravitating Fermi gas giving the temperature as a function of
energy and investigate the nature of phase transitions as a function
of the dimension of space. We consider stable states (global entropy
maxima) as well as metastable states (local entropy maxima). We show
that for $D\ge 4$, there exists a critical temperature (for
sufficiently large systems) and a critical energy below which the
system cannot be found in statistical equilibrium. Therefore, for
$D\ge 4$, quantum mechanics cannot stabilize matter against
gravitational collapse. This is similar to a result found by Ehrenfest
(1917) at the atomic level for Coulombian forces. This makes the
dimension $D=3$ of our universe very particular with possible
implications regarding the anthropic principle. Our study enters in a
long tradition of scientific and philosophical papers who studied how
the dimension of space affects the laws of physics.

\end{abstract}

\section{Introduction}
\label{sec_introduction}

The statistical mechanics of systems with long-range interactions is
currently a topic of active research \cite{dauxois}. Among long-range
interactions, the gravitational force plays a fundamental
role. Therefore, the developement of a statistical mechanics for
self-gravitating systems is of considerable interest \cite{paddy}. In
this context, a system of self-gravitating fermions enclosed within a
box provides an interesting model which can be studied in great detail
\cite{ht,pt}. This model incorporates an effective small-scale cut-off
played by the Pauli exclusion principle and a large scale cut-off
played by the confining box (other forms of confinement could also
be considered). The statistical mechanics of this system is rigorously
justified and presents a lot of interesting features which are of
interest in statistical mechanics \cite{houches} and astrophysics
\cite{chavcape}. Its detailed study is therefore important at a
conceptual and practical level.

In a preceding paper \cite{pt}, we have discussed the nature of phase
transitions in the self-gravitating Fermi gas 
in a space of dimension $D=3$. Our study was performed in both
microcanonical and canonical ensembles and considered an arbitrary
degree of degeneracy relative to the system size. This study completes
previous investigations by Hertel \& Thirring \cite{ht} who worked in
the canonical ensemble and considered small system sizes. At high
temperatures and high energies, the system is in a gaseous phase and
quantum effects are completely negligible. At some transition
temperature $T_{t}$ or transition energy $E_{t}$ (for sufficiently
large system sizes), a first order phase transition is expected to
occur and drive the system towards a condensed phase.  However,
gaseous states are still metastable below this transition point and
gravitational collapse will rather occur at a smaller critical
temperature $T_{c}$ (Jeans temperature) \cite{aa} or critical energy
$E_{c}$ (Antonov energy) \cite{antonov,lbw,paddy} at which the
metastable branch disappears (spinodal point). The end-state of the
collapse is a compact object with a ``core-halo''
structure. Typically, it consists of a degenerate nucleus surrounded
by a ``vapour''. The nucleus (condensate) resembles a white dwarf star
\cite{chandra2}. At non-zero temperature, this compact object is
surrounded by a dilute atmosphere. Therefore, when quantum mechanics
is properly accounted for, there exists an equilibrium state (global
maximum of entropy or free energy) for each value of accessible energy and
temperature.  The condensate results from the balance between
gravitational contraction and quantum pressure. As first noticed by
Fowler \cite{fowler} in his classical theory of white dwarf stars,
quantum mechanics is able to stabilize matter against gravitational
collapse.

One object of this paper is to show that this conclusion is no more
valid in a space of dimension $D\ge 4$. For a system of mass $M$
enclosed within a box of radius $R$, there exists a critical
temperature (for sufficiently large $R$) and a critical
energy below which the system cannot be found at statistical
equilibrium. This is like the Antonov instability for self-gravitating
classical particles in $D=3$
\cite{antonov,lbw,paddy} but it now occurs for
fermions. Therefore, quantum mechanics cannot arrest gravitational
collapse in $D\ge 4$. This result is connected to our previous
observation \cite{langevin} that a classical white dwarf star (a
polytrope of index $n_{3/2}=D/2$) becomes unstable for $D\ge
4$. Interestingly, this result is similar to that of Ehrenfest
\cite{ehrenfest} who considered the stability of atomic structures
(in Bohr's model) for different dimensions of space and concludes that
$D<4$ is required for stability. In this paper, we determine the
caloric curve of the self-gravitating Fermi gas for an arbitrary
dimension of space and an arbitrary degree of degeneracy (or system
size).  We exhibit particular dimensions that play a special role in
the problem. The dimension $D=2$ is critical because the results
established for $D\neq 2$ cannot be directly extended to $D=2$
\cite{sc1}.  Furthermore, in $D=2$ the radius of a white dwarf star is
independent on its mass and given in terms of fundamental constants by
$R=0.27 \ h m^{-3/2} G^{-1/2}$. The dimension $D=4$ is also critical
because it is the dimension at which classical white dwarf stars
become unstable. At this particular dimension, their mass is
independent on radius and can be expressed in terms of fundamental
constants as $M=1.44\ 10^{-2}h^{4}m^{-5}G^{-2}$. Mathematically, this
is similar to Chandrasekhar's limiting mass \cite{chandra1} for
relativistic white dwarf stars in $D=3$.  The dimension
$D=2(1+\sqrt{2})$ is also particular because at this dimension, the
white dwarf stars cease to be self-confined and have infinite
mass. Finally, $D=10$ is the dimension at which the caloric curve of
classical isothermal spheres loses its characteristic spiral nature
\cite{sc1}. Although we systematically explore all dimensions of space 
in order to have a complete picture of the problem, only dimensions
$D=1$, $D=2$ and $D=3$ are {\it a priori} of physical interest. The
dimension $D=1$ is considered in cosmology and in connexion with shell
models, and the dimension $D=2$ can be useful to describe filaments or
ring structures with high aspect ratio. Two-dimensional gravity is
also of interest for its properties of conformal invariance and for
its relation with two-dimensional turbulence \cite{houches}. Non-integer
dimensions can arise if the system has a fractal nature.

The paper is organized as follows. In Sec. \ref{sec_fermions}, we
determine the thermodynamical parameters of the self-gravitating Fermi
gas in dimension $D$. The Fermi-Dirac entropy is introduced from a
combinatorial analysis. In Sec. \ref{sec_lim}, we consider asymptotic
limits corresponding to the classical self-gravitating gas and to
completely degenerate structures (white dwarfs). We emphasize the
importance of metastable states in astrophysics and explain how they
can be taken into account in the theory (see also \cite{meta}). We
also discuss the thermodynamic limit of the self-gravitating quantum
gas and compare it with the thermodynamic limit of the
self-gravitating classical gas in the dilute limit \cite{vs}. In
Sec. \ref{sec_caloric}, we provide a gallery of caloric curves of the
self-gravitating Fermi gas in different dimensions of space.  Rigorous
mathematical results on the existence of solutions of the
Fermi-Poisson equation have been obtained by Sta\'nczy
\cite{stanczy}. Finally, in the conclusion, we place our study in a
more general perspective. We give a short historical account of
scientific and philosophical papers who studied the role played by the
dimension of space in determining the form of the laws of
physics. These works tend to indicate that the dimension $D=3$ of our
universe is very particular. This is also the result that we reach in
our study. These remarks can have implications regarding the anthropic
principle.

\section{Thermodynamics of self-gravitating $D$-fermions}
\label{sec_fermions}

\subsection{The Fermi-Dirac distribution}
\label{sec_fd}

We consider a system of $N$ fermions interacting via Newtonian gravity
in a space of dimension $D$. We assume that the mass of the
configuration is sufficiently small so as to ignore relativistic
effects. Let $f({\bf r},{\bf v},t)$ denote the distribution function
of the system, i.e. $f({\bf r},{\bf v},t)d^{D}{\bf r} d^{D}{\bf v}$
gives the mass of particles whose position and velocity are in the
cell $({\bf r},{\bf v};{\bf r}+d^{D}{\bf r},{\bf v}+d^{D}{\bf v})$ at
time $t$. The integral of $f$ over the velocity determines the spatial
density
\begin{equation}
\rho=\int f d^{D}{\bf v},
\label{fd1}
\end{equation}
and the total mass of the configuration is given by
\begin{equation}
M=\int \rho d^{D}{\bf r},
\label{fd2}
\end{equation}
where the integral extends over the entire domain. On the other hand,
in the meanfield approximation, the total energy of the system can be
expressed as
\begin{equation}
E={1\over 2}\int fv^{2}d^{D}{\bf r}d^{D}{\bf v}+{1\over 2}\int\rho\Phi d^{D}{\bf r}=K+W,
\label{fd3}
\end{equation}
where $K$ is the kinetic energy and $W$ the potential energy. The
gravitational potential $\Phi$ is related to the density by the
Newton-Poisson equation
\begin{equation}
\Delta\Phi=S_{D} G\rho,
\label{fd4}
\end{equation}
where $S_{D}=2\pi^{D/2}/\Gamma(D/2)$ is the surface of a unit
sphere in a space of dimension $D$ and $G$ is the constant of gravity
(which depends on the dimension of space).

We now wish to determine the most probable distribution of
self-gravitating fermions at statistical equilibrium. To that
purpose, we divide the individual phase space $\lbrace {\bf
r},{\bf v}\rbrace$ into a very large number of microcells with
size $(h/m)^D$ where $h$ is the Planck constant (the mass $m$ of
the particles arises because we use ${\bf v}$ instead of ${\bf p}$
as a phase space coordinate). A microcell is occupied either by $0$
or $1$ fermion (or $g=2s+1$ fermions if we account for the spin). We shall
now group these microcells into macrocells each of which contains
many microcells but remains nevertheless small compared to the
phase-space extension of the whole system. We call $\nu$ the
number of microcells in a macrocell. Consider the configuration
$\lbrace n_i \rbrace$ where there are $n_1$ fermions in the
$1^{\rm st}$ macrocell, $n_2$ in the $2^{\rm nd}$ macrocell
etc..., each occupying one of the $\nu$ microcells with no
cohabitation. The number of ways of assigning a microcell to the
first element of a macrocell is $\nu$, to the second $\nu -1$ etc.
Since the particles are indistinguishable, the number of ways of
assigning microcells to all $n_i$ particles in a macrocell is thus
\begin{equation}
{1\over n_i!}{\times} {\nu!\over (\nu-n_i)!}. \label{fd5}
\end{equation}
To obtain the number of microstates corresponding to the macrostate
$\lbrace n_i \rbrace$ defined by the number of fermions $n_i$ in each
macrocell (irrespective of their precise position in the cell), we
need to take the product of terms such as (\ref{fd5}) over all
macrocells. Thus, the number of microstates corresponding to the
macrostate $\lbrace n_i \rbrace$, i.e. the probability of the state
$\lbrace n_i \rbrace$, is
\begin{equation}
W(\lbrace n_i \rbrace)=\prod_i {\nu!\over n_i!(\nu-n_i)!}.
\label{fd6}
\end{equation}
This is the Fermi-Dirac statistics. As is customary, we define the
entropy of the state $\lbrace n_i \rbrace$ by
\begin{equation}
S(\lbrace n_i \rbrace)=\ln W(\lbrace n_i \rbrace). \label{fd7}
\end{equation}
It is convenient here to return to a representation in terms of
the distribution function giving the phase-space density in the
$i$-th macrocell
\begin{equation}
f_i=f({\bf r}_i,{\bf v}_i)={n_i \ m\over \nu \ ({h\over
m})^D}={n_i\eta_0\over \nu}, \label{fd8}
\end{equation}
where we have defined $\eta_0=m^{D+1}/h^D$, which represents the
maximum value of $f$ due to Pauli's exclusion principle. Now,
using the Stirling formula, we have
\begin{equation}
\ln W(\lbrace n_i \rbrace)\simeq \sum_i \nu
(\ln\nu-1)-\nu\biggl\lbrace {f_i\over \eta_0}\biggl\lbrack
\ln\biggl ({\nu f_i\over \eta_0}\biggr )-1\biggr\rbrack +\biggl
(1-{f_i\over \eta_0}\biggr )\biggl\lbrack\ln\biggl\lbrace
\nu\biggl (1-{f_i\over \eta_0}\biggr
)\biggr\rbrace-1\biggr\rbrack\biggr\rbrace. \label{fd9}
\end{equation}
Passing to the continuum limit $\nu\rightarrow 0$, we obtain the
usual expression of the Fermi-Dirac entropy
\begin{equation}
S=-k_B\int \biggl\lbrace {f\over\eta_{0}}\ln
{f\over\eta_{0}}+\biggl (1-  {f\over\eta_{0}}\biggr)\ln \biggl (1-
{f\over\eta_{0}}\biggr)\biggr\rbrace\ {d^{D}{\bf r}d^{D}{\bf
v}\over ({h\over m})^D}. \label{fd10}
\end{equation}
If we take into account the spin of the particles, the above
expression remains valid but the maximum value of the distribution
function is now  $\eta_{0}=g m^{D+1}/h^{D}$, where $g=2s+1$ is the
spin multiplicity of the quantum states (the phase space element
has also to be multiplied by $g$). An expression of entropy
similar to (\ref{fd10}), but arising for a completely different reason,
has been introduced by Lynden-Bell in the context of the violent
relaxation of collisionless stellar systems
\cite{lb,cs,dubrovnik}. In that context, $\eta_0$ represents the
maximum value of the initial distribution function and the actual
distribution function (coarse-grained) must always satisfy
$\overline{f}\le \eta_0$  by virtue of the Liouville theorem. This
is the origin of the ``effective'' exclusion principle in
Lynden-Bell's theory, which has nothing to do with quantum
mechanics. Since the particles (stars) are distinguishable
classical objects (but subject to an exclusion principle in the
collisionless regime), Lynden-Bell's statistics corresponds to a
$4$-th form of statistics (in addition to the Maxwell-Boltzmann,
Fermi-Dirac and Bose-Einstein statistics). However, for a single type of
phase element $\eta_0$, it leads to the same results as the
Fermi-Dirac statistics. We also recall that in the non-degenerate
(or classical) limit $f\ll\eta_0$, the Fermi-Dirac entropy (\ref{fd10})
reduces to the Boltzmann entropy
\begin{equation}
S=-k_B\int {f\over m}\biggl\lbrack \ln\biggl ({f h^D\over g
m^{D+1}}\biggr )-1\biggr \rbrack d^{D}{\bf r}d^{D}{\bf v}.
\label{fd11}
\end{equation}

Now that the entropy has been precisely justified, the statistical
equilibrium state (most probable state) of self-gravitating
fermions is obtained by maximizing the Fermi-Dirac entropy (\ref{fd10}) at
fixed mass (\ref{fd2}) and energy (\ref{fd3}):
\begin{equation}
{\rm Max}\quad S[f]\quad | \quad E[f]=E, M[f]=M. \label{fd12}
\end{equation}
Introducing Lagrange multipliers
$1/T$ (inverse temperature) and $\mu$ (chemical potential)
to satisfy these constraints, and writing the variational
principle in the form
\begin{equation}
\delta S-{1\over T}\ \delta E+{\mu\over T} \delta N=0, \label{fd13}
\end{equation}
we find that the {\it critical points} of entropy
correspond to the Fermi-Dirac distribution
\begin{equation}
f={\eta_{0}\over 1+\lambda e^{\beta m ({v^{2}\over 2}+\Phi)}},
\label{fd14}
\end{equation}
where $\lambda=e^{-\beta \mu}$ is a strictly positive constant
(inverse fugacity) and $\beta={1\over k_B T}$ is the inverse
temperature. Clearly, the
distribution function satisfies $f\le \eta_{0}$, which is a
consequence of Pauli's exclusion principle.

So far, we have assumed that the system is isolated so that the energy
is conserved. If now the system is in contact with a thermal bath
(e.g., a radiation background) fixing the temperature, the statistical
equilibrium state minimizes the free energy $F=E-TS$, or maximizes the
Massieu function $J=S-\beta E$, at fixed mass and temperature:
\begin{equation}
{\rm Max}\quad J[f]\quad |\quad M[f]=M. \label{fd15}
\end{equation}
Introducing Lagrange multipliers and writing the variational
principle in the form
\begin{equation}
\delta J+{\mu\over T} \delta N=0, \label{fd16}
\end{equation}
we find that the {\it critical points} of free energy are again given by
the Fermi-Dirac distribution (\ref{fd14}). Therefore, the critical
points (first variations) of the variational problems (\ref{fd12}) and
(\ref{fd15}) are the same. However, the stability of the system
(regarding the second variations) can be different in microcanonical
and canonical ensembles. When this happens, we speak of a situation of
{\it ensemble inequivalence} \cite{pt}. The stability of the system can be
determined by a graphical construction, by simply plotting the caloric
curve/series of equilibria $\beta(E)$ and using the turning
point method of Katz \cite{katz,katz2}.

\subsection{Thermodynamical parameters}
\label{sec_para}

Integrating the distribution function (\ref{fd14}) over velocity, we find that
the density of particles is related to the gravitational potential by
\begin{equation}
\rho={\eta_{0}S_{D}2^{{D/2}-1}\over (\beta m)^{D/2}}I_{D/2-1}(\lambda e^{\beta m\Phi}),
\label{p1}
\end{equation}
where $I_{n}$ denotes the Fermi integral
\begin{equation}
I_{n}(t)=\int_{0}^{+\infty}{x^{n}\over 1+t e^{x}}dx.
\label{p2}
\end{equation}
We recall the identity
\begin{equation}
I'_{n}(t)=-{n\over t}I_{n-1}(t), \qquad (n>0),
\label{p3}
\end{equation}
which can be established from (\ref{p2}) by an integration by parts. The
gravitational potential is now obtained by substituting Eq. (\ref{p1}) in the
Poisson equation (\ref{fd4}). We introduce the rescaled distance $\xi=\lbrack
{S_{D}^{2} 2^{D/2-1}G\eta_{0}/(\beta m)^{D/2-1}}\rbrack^{1/2}r$ and the
variables $\psi=\beta m (\Phi-\Phi_{0})$ and $k=\lambda
e^{\beta m \Phi_{0}}$, where $\Phi_{0}$ is the central potential. Thus,
we get the $D$-dimensional Fermi-Poisson equation
\begin{equation}
{1\over\xi^{D-1}}{d\over d\xi}\biggl (\xi^{D-1}{d\psi\over d\xi}\biggr )=I_{D/2-1}(ke^{\psi(\xi)}),
\label{p4}
\end{equation}
\begin{equation}
\psi(0)=\psi'(0)=0.
\label{p5}
\end{equation}

As is well-known, self-gravitating systems at non-zero temperature
have the tendency to evaporate. Therefore, there is no equilibrium
state in a strict sense and the statistical mechanics of
self-gravitating systems is essentially an out-of-equilibrium
problem. However, the evaporation rate is small in general and the
system can be found in a quasi-equilibrium state for a relatively long
time.  In order to describe the thermodynamics of the self-gravitating
Fermi gas rigorously, we shall use an artifice and enclose the system
within a spherical box of radius $R$ (the box typically represents the
size of the cluster under consideration). In that case, the solution
of Eq. (\ref{p4}) is terminated by the box at the normalized radius
\begin{equation}
\alpha=\biggl \lbrack {S_{D}^{2} 2^{D/2-1} G\eta_{0}\over (\beta m)^{D/2-1}}\biggr \rbrack^{1/2} R.
\label{p6}
\end{equation}
For a spherically symmetric configuration, the Gauss theorem can be
written
\begin{equation}
{d\Phi\over dr}={GM(r)\over r^{D-1}},
\label{p7}
\end{equation}
where $M(r)=\int_{0}^{r}\rho S_{D}r^{D-1}dr$ is the mass within the sphere of radius $r$. Applying this result at $r=R$ and using the variables introduced previously we get
\begin{equation}
\eta\equiv {\beta GMm\over R^{D-2}}=\alpha\psi'_{k}(\alpha).
\label{p8}
\end{equation}
This equation relates the dimensionless box radius $\alpha$ and the uniformizing variable $k$ to the dimensionless inverse temperature $\eta$. According to Eqs. (\ref{p6}) and (\ref{p8}), $\alpha$ and $k$ are related to each other by the relation $\alpha^{2}\eta^{D/2-1}=\mu$ or, explicitly,
\begin{equation}
\alpha^{D+2\over D-2}\psi'_{k}(\alpha)=\mu^{2\over D-2},
\label{p9}
\end{equation}
where 
\begin{equation}
\mu=\eta_{0}\sqrt{S_{D}^{4}2^{D-2}G^{D}M^{D-2}R^{D(4-D)}},
\label{p10}
\end{equation}
is the degeneracy parameter \cite{cs}. It should not be confused with
the chemical potential. We shall give a physical interpretation of
this parameter in Sec. \ref{sec_d3}.

The calculation of the energy is a little more involved. First, we introduce the local pressure
\begin{equation}
p={1\over D}\int f v^{2}d^{D}{\bf v}.
\label{p11}
\end{equation}
Using the Fermi-Dirac distribution function (\ref{fd14}), we find that
\begin{equation}
p={\eta_{0}S_{D}2^{{D/2}}\over D(\beta m)^{D/2+1}}I_{D/2}(k e^{\psi}).
\label{p12}
\end{equation}
The kinetic energy $K=(D/2)\int p d^{D}{\bf r}$ can thus be written
\begin{equation}
{K R^{D-2}\over GM^{2}}={\alpha^{4+4D-D^{2}\over D-2}\over \mu^{4\over D-2}}\int_{0}^{\alpha}I_{D/2}(ke^{\psi_{k}(\xi)})\xi^{D-1}d\xi.
\label{p13}
\end{equation}
In order to determine the potential energy, we use the $D$-dimensional version of the Virial theorem \cite{langevin}. For $D\neq 2$, it reads
\begin{equation}
2K+(D-2)W=DV_{D}R^{D}p(R),
\label{p14}
\end{equation}
where $V_{D}=S_{D}/D$ is the volume of a hypersphere with unit radius (the case $D=2$ will be considered specifically in Sec. \ref{sec_d2}). Using the expression of the pressure (\ref{p12}) at the box radius $R$, we get
\begin{equation}
{W R^{D-2}\over GM^{2}}={2\over D(D-2)}{\alpha^{2(D+2)\over D-2}\over \mu^{4\over D-2}}I_{D/2}(k e^{\psi(\alpha)})-{2KR^{D-2}\over (D-2)GM^{2}}.
\label{p15}
\end{equation}
Combining Eqs. (\ref{p13}) and (\ref{p15}), we finally obtain
\begin{equation}
\Lambda\equiv -{E R^{D-2}\over GM^{2}}={4-D\over D-2}{\alpha^{4+4D-D^{2}\over D-2}\over \mu^{4\over D-2}}\int_{0}^{\alpha}I_{D/2}(ke^{\psi_{k}(\xi)})\xi^{D-1}d\xi-{2\over D(D-2)}{\alpha^{2(D+2)\over D-2}\over \mu^{4\over D-2}}I_{D/2}(k e^{\psi(\alpha)}).
\label{p16}
\end{equation}
For $D=3$, Eqs. (\ref{p8}) and (\ref{p16}) return the expressions
derived in \cite{cs,pt}. For a given value of $\mu$ and $k$, we can
solve the ordinary differential equation (\ref{p4}) until the value of
$\alpha$ at which the condition (\ref{p9}) is satisfied. Then,
Eqs. (\ref{p8}) and (\ref{p16}) determine the temperature and the
energy of the configuration. By varying the parameter $k$ (for a fixed
value of the degeneracy parameter $\mu$), we can determine the full
caloric curve/series of equilibria $\beta(E)$. Extending the
results of \cite{pt} in $D$ dimensions, the entropy of each
configuration, parameterized by $\alpha$, is given by
\begin{equation}
{S\over Nk_{B}}=-{4+4D-D^{2}\over D(4-D)}\Lambda\eta+\psi_{k}(\alpha)+{\eta\over D-2}+\ln k-{2(D-2)\over D^{2}(4-D)}{\alpha^{2D\over D-2}\over \mu^{2\over D-2}}I_{D/2}(ke^{\psi_{k}(\alpha)}),
\label{p17}
\end{equation}
and the free energy by
\begin{equation}
F=E-TS.
\label{p18}
\end{equation}
In the microcanonical ensemble, a solution is stable if it corresponds
to a maximum of entropy $S[f]$ at fixed mass and energy.  In the
canonical ensemble, the condition of stability requires that the
solution be a minimum of free energy $F[f]$ at fixed mass and
temperature. This meanfield approach is {\it exact} in a
thermodynamical limit such that $N\rightarrow +\infty$ with $\mu$,
$\eta$, $\Lambda$ fixed. If we fix $\eta_{0}$ (i.e. $\hbar$) and $G$,
this implies that $RN^{(D-2)/(D(4-D))}$, $TN^{-4/(D(4-D))}$,
$EN^{-(4D-D^{2}+4)/(D(4-D))}$, $SN^{-1}$ and $JN^{-1}$ approach a
constant value for $N\rightarrow +\infty$ (the free energy $F$ scales
as $N^{(4D-D^{2}+4)/(D(4-D))}$). This is the quantum thermodynamic
limit (QTL) for the self-gravitating gas \cite{pt,rieutord}. The usual
thermodynamic limit $N,R\rightarrow +\infty$ with $N/R^{D}$ constant
is clearly not relevant for inhomogeneous systems whose energy is non-additive.

\section{Asymptotic limits}
\label{sec_lim}

\subsection{The non degenerate limit $(\mu=\infty)$}
\label{sec_class}

Before considering the case of an arbitrary degree of degeneracy, it
may be useful to discuss first the non degenerate limit corresponding
to a classical isothermal gas ($\hbar\rightarrow 0$). For
$f\ll\eta_{0}$, the distribution function (\ref{fd14}) reduces to the
Maxwell-Boltzmann formula
\begin{equation}
f={\eta_{0}\over\lambda}e^{-\beta m({v^{2}\over 2}+\Phi)},
\label{cc1}
\end{equation}
which can be written more conveniently as
\begin{equation}
f=\biggl ({\beta m\over 2\pi}\biggr )^{D/2}\rho({\bf r})\ e^{-\beta m {v^{2}\over 2}}.
\label{cc1bis}
\end{equation}
The density profile can be written
\begin{equation}
\rho=\rho_{0}e^{-\psi(\xi)}, \label{cc2}
\end{equation}
where $\rho_{0}$ is the central density, $\xi$ is the normalized distance
\begin{equation}
\xi=(S_{D} G\beta m\rho_{0})^{1/2}r,
\label{cc3}
\end{equation}
and $\psi$ is the solution of the $D$-dimensional Emden equation
\begin{equation}
{1\over \xi^{D-1}}{d\over d\xi}\biggl (\xi^{D-1}{d\psi\over d\xi}\biggr )=e^{-\psi},
\label{cc4}
\end{equation}
with boundary conditions
\begin{equation}
\psi(0)=\psi'(0)=0.
\label{cc5}
\end{equation}
This equation can be obtained from Eq. (\ref{p4}) by taking the limit $k\rightarrow +\infty$ and using the limiting form of the Fermi integral
\begin{equation}
I_{n}(t)\sim {1\over t}\Gamma(n+1), \qquad (t\rightarrow +\infty).
\label{cc6}
\end{equation}
From Eq. (\ref{cc1bis}), we check that the local equation of state of a
classical self-gravitating isothermal gas is $p({\bf r})={\rho({\bf r})\over
m}k_{B}T$ whatever the dimension of space. The thermodynamical
parameters are given by
\begin{equation}
\eta=\alpha\psi'(\alpha),
\label{cc7}
\end{equation}
\begin{equation}
\Lambda={D(4-D)\over 2(D-2)}
{1\over\alpha\psi'(\alpha)}-{1\over
D-2}{e^{-\psi(\alpha)}\over\psi'(\alpha)^{2}},
\label{cc8}
\end{equation}
\begin{equation}
{S-S_{0}\over Nk_{B}}=-{D-2\over 2}\ln\eta-2\ln\alpha+\psi(\alpha)+{\eta\over D-2}-2\Lambda\eta,
\label{cc9}
\end{equation}
\begin{equation}
{S_{0}\over Nk_{B}}=\ln\mu+\ln\biggl ({2\pi^{D/2}\over S_{D}}\biggr )+1-{D\over 2},
\label{cc9bis}
\end{equation}
where $\alpha=(S_{D} G\beta m\rho_{0})^{1/2}R$ is the normalized box
radius. For $D=2$, the thermodynamical parameters can be calculated
analytically \cite{sc1}. Introducing the pressure at the box $P=p(R)$,
the global equation of state of the self-gravitating gas can be written
\begin{equation}
{PV\over Nk_{B}T}={1\over D}{\alpha^{2}\over \eta}e^{-\psi(\alpha)}.
\label{cc11}
\end{equation}
We recall that the foregoing expressions can be expressed in terms of
the value of the Milne variables $u_{0}=u(\alpha)$ and
$v_{0}=v(\alpha)$ at the normalized box radius \cite{aa,grand}. The
structure and the stability of classical isothermal spheres in $D$
dimensions have been studied in detail in \cite{sc1}. The classical
thermodynamic limit (CTL) of self-gravitating systems, or dilute limit
\cite{vs}, is such that $N\rightarrow +\infty$ with $\eta$, $\Lambda$
fixed. If we take $\beta\sim 1$, this implies that $R\sim N^{1/(D-2)}$
and $E,S,J,F\sim N$. The physical distinction between the quantum
thermodynamic limit (QTL) and the classical thermodynamic limit (CTL)
is related to the existence of long-lived gaseous metastable states as
discussed in
\cite{rieutord,meta}.

\subsection{The completely degenerate limit}
\label{sec_deg}

For $\beta\rightarrow +\infty$ (i.e., $T=0$), the distribution
function (\ref{fd14}) reduces to a step function: $f=\eta_{0}$ if
$v\le v_{F}$ and $f=0$ if $v\ge v_{F}$, where $v_{F}({\bf
r})=\sqrt{2(\mu/m-\Phi)}$ is the local Fermi velocity. In that case,
the density and the pressure can be explicitly evaluated:
\begin{equation}
\rho=\int_{0}^{v_{F}}\eta_{0}S_{D}v^{D-1}dv=\eta_{0}S_{D}{v_{F}^{D}\over D},
\label{d1}
\end{equation}
\begin{equation}
p={1\over D}\int_{0}^{v_{F}}\eta_{0}S_{D}v^{D+1}dv=\eta_{0}{S_{D}\over D}{v_{F}^{D+2}\over D+2}.
\label{d2}
\end{equation}
Eliminating the Fermi velocity between these two expressions, we find that  the
equation of state of a cold Fermi gas in $D$ dimensions is
\begin{equation}
p=K\rho^{1+2/D}, \qquad K={1\over D+2}\biggl ({D\over\eta_{0}S_{D}}\biggr )^{2/D}.
\label{d3}
\end{equation}
This equation of state describes a $D$-dimensional classical white
dwarf star (throughout this paper, we shall call  ``white dwarf
star'', or ``fermion ball'', a completely degenerate self-gravitating
system. This terminology will be extended to any dimension of
space).  In $D=3$, classical white dwarf stars are equivalent to
polytropes with index $n=3/2$ \cite{fowler}.  In $D$ dimensions,
classical ``white dwarf stars'' are equivalent to polytropes with
index \cite{langevin}:
\begin{equation}
n_{3/2}={D\over 2}.
\label{d4}
\end{equation}
The structure and the stability of polytropic spheres in $D$
dimensions have been studied in detail in \cite{langevin}. It is shown
that a polytrope of index $n$ is self-confined for
$n<n_{5}=(D+2)/(D-2)$ and stable for $n<n_{3}=D/(D-2)$. Therefore,
white dwarf stars ($n=n_{3/2}=D/2$) are self-confined only for
$D<2(1+\sqrt{2})$ and they are stable only for $D\le 4$. For $D>4$,
quantum mechanics is not able to stabilize matter against
gravitational collapse. Thus, $D=4$ is a critical dimension regarding
gravitational collapse.  $D=2$ is also critical \cite{sc1}. Therefore,
the dimension of space of our universe $2<D=3<4$ lies between two
critical dimensions.

\begin{figure}[htbp]
\centerline{
\psfig{figure=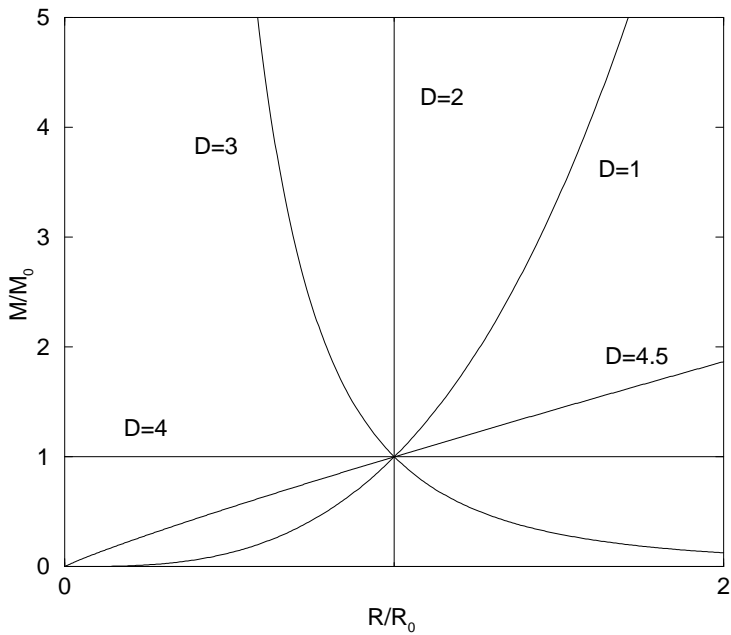,angle=0,height=8.5cm}}
\caption{The mass-radius relation for complete white dwarf stars ($T=0$) in different dimensions of space. It clearly shows that the dimension $D=3$ in surrounded by two critical dimensions $D=2$ and $D=4$ at which either the radius or the mass is constant.}
\label{MRD}
\end{figure}

We now introduce dimensionless parameters associated with $n_{3/2}$
polytropes which will be useful in the sequel. Their density profile can
be written
\begin{equation}
\rho(r)=\rho_{0}\ \theta^{D/2}(\xi),
\label{d5}
\end{equation}
where $\rho_{0}$ is the central
density, $\xi$ is the normalized distance
\begin{equation}
\xi=\biggl \lbrack {2S_{D}G\rho_{0}^{(D-2)\over D}\over
K(D+2)}\biggr \rbrack^{1/2} r,
\label{d6}
\end{equation}
and $\theta$ is solution of the $D$-dimensional Lane-Emden
equation
\begin{equation}
{1\over \xi^{D-1}}{d\over d\xi}\biggl (\xi^{D-1}{d\theta\over d\xi}\biggr )=-\theta^{D/2},
\label{d7}
\end{equation}
with boundary conditions
\begin{equation}
\theta(0)=1, \qquad \theta'(0)=0.
\label{d8}
\end{equation}
This equation can be obtained from Eq. (\ref{p4}) by taking the limit
$k\rightarrow 0$ and using the limiting form of the Fermi integral
\begin{equation}
I_{n}(t)\sim {(-\ln t)^{n+1}\over n+1}, \qquad (t\rightarrow 0).
\label{d9}
\end{equation}
For $D<2(1+\sqrt{2})$, the solution of the Lane-Emden equation (\ref{d7})
vanishes at a finite distance $\xi_{1}$ defining the radius $R_{*}$ of
the white dwarf star ({\it complete polytrope}). Using the results of
\cite{langevin}, the mass-radius relation of $D$-dimensional white
dwarf stars is given by
\begin{equation}
M^{D-2\over D}R_{*}^{4-D}={K(D+2)\over 2 G S_{D}^{2/D}}\omega_{D/2}^{D-2\over D},
\label{d10}
\end{equation}
where we have defined
\begin{equation}
\omega_{D/2}=-\xi_{1}^{D+2\over D-2}\theta'(\xi_{1}).
\label{d11}
\end{equation}
For $2<D<4$, the mass $M$ decreases with the radius $R_{*}$ while for
$D<2$ and for $4<D<2(1+\sqrt{2})$ it increases with the radius (see
Fig. \ref{MRD}). The mass-radius relation (\ref{d10}) exhibits the two
critical dimensions of space $D=2$ and $D=4$ discussed previously. For
$D=2$, the radius is independent on mass and for $D=4$, the mass is
independent on radius (see Sec. \ref{sec_caloric}). The energy of a
self-confined white dwarf star is
\begin{equation}
E=-\lambda_{D/2}{ GM^{2}\over R_{*}^{D-2}},
\label{d14}
\end{equation}
where
\begin{equation}
\lambda_{D/2}={D(4-D)\over (D-2)(4+4D-D^{2})}.
\label{d15}
\end{equation}
We note that the energy of a white dwarf star vanishes for $D=4$.
According to Poincar\'e's theorem \cite{chandra2}, this determines the
onset of instability. We thus recover the
fact that complete white dwarf stars are unstable for $D>4$ \cite{langevin}.

For $D>2(1+\sqrt{2})$, the density of a $n_{3/2}$ polytrope never
vanishes (as $n_{3/2}>n_{5}$) and we need to confine the system within
a box of radius $R$ ({\it incomplete polytrope}) to avoid the infinite
mass problem. In that case, the white dwarf star exerts a pressure
against the box. White dwarf stars with $R_{*}>R$ when
$D<2(1+\sqrt{2})$ are also incomplete. They are arrested by the box at
the normalized radius $\xi=\alpha$ with $\alpha=\lbrace
2S_{D}G\rho_{0}^{(D-2)/D}/\lbrack K(D+2)\rbrack\rbrace^{1/2}R$. As
shown in \cite{langevin}, the normalized mass and the normalized
energy of the configuration parameterized by $\alpha$ are given by
\begin{equation}
\eta_{P}\equiv {M\over S_{D}}\biggl\lbrack {2S_{D}G\over K(D+2)}\biggr \rbrack^{D\over D-2}{1\over R^{D(D-4)\over D-2}}=-\alpha^{D+2\over D-2}\theta'(\alpha),
\label{d17}
\end{equation}
\begin{equation}
\Lambda\equiv -{ER^{D-2}\over GM^{2}}={-2\over D^{2}-4D-4}\biggl\lbrace {D(4-D)\over 2(D-2)}\biggl \lbrack 1+(D-2){\theta(\alpha)\over\alpha\theta'(\alpha)}\biggr \rbrack+{2-D\over 2+D}{\theta(\alpha)^{D+2\over 2}\over \theta'(\alpha)^{2}}\biggr\rbrace.
\label{d18}
\end{equation}
In the present context, the normalized mass $\eta_P$ is related to
the degeneracy parameter $\mu$ by the relation
\begin{equation}
\eta_{P}=\biggl ({2\mu\over D}\biggr )^{2\over D-2}. \label{d19}
\end{equation}
On the other hand, using Eqs. (\ref{d10}) and (\ref{d14}), the
normalized mass and the normalized energy of a self-confined white
dwarf star with $R_{*}<R$ (complete polytrope) are given by
\begin{equation}
\eta_{P}=\omega_{D/2}\biggl ({R_{*}\over R}\biggr )^{(D-4)D\over
D-2} \label{d20}
\end{equation}
\begin{equation}
\Lambda=\lambda_{D/2}\biggl ({R\over R_{*}}\biggr )^{D-2}.
\label{d21}
\end{equation}
Eliminating $R_{*}$ between these two relations, we obtain the
``mass-energy'' relation
\begin{equation}
\Lambda \eta_{P}^{(D-2)^{2}\over D(D-4)}=\lambda_{D/2}
(\omega_{D/2})^{(D-2)^{2}\over D(D-4)}, \label{d22}
\end{equation}
which will be useful in our subsequent analysis.

\section{Caloric curves in various dimensions}
\label{sec_caloric}

\subsection{Series of equilibria and metastable states} 
\label{sec_seq}

We shall now determine the caloric curve $\beta(E)$ of the
self-gravitating Fermi gas as a function of the degeneracy parameter
$\mu$ for any dimension of space $D$. This study has already been
performed for $D=3$ in \cite{pt}. The critical points of the
Fermi-Dirac entropy $S[f]$ at fixed $E$ and $M$ (i.e., the
distribution functions $f({\bf r},{\bf v})$ which cancel the first
order variations of $S$ at fixed $E$, $M$) form a series of equilibria
parameterized by the uniformizing variable $k$. At each point in the
series of equilibria corresponds a temperature $\beta$ and an energy
$E$ determined by Eqs. (\ref{p8}) and (\ref{p16}).  In this approach,
$\beta$ is the Lagrange multiplier associated with the conservation of
energy in the variational problem (\ref{fd13}). It has also the
interpretation of a kinetic temperature in the Fermi-Dirac
distribution (\ref{fd14}). We can thus plot $\beta(E)$ along the
series of equilibria. There can be several values of temperature
$\beta$ for the same energy $E$ because the variational problem
(\ref{fd12}) can have several solutions: a local entropy maximum
(metastable state), a global entropy maximum, and one or several
saddle points. We shall represent all these solutions on the caloric
curve because local entropy maxima (metastable states) are in general
more physical than global entropy maxima for the timescales achieved
in astrophysics. Indeed, the system can remain frozen in a metastable
gaseous phase for a very long time. This is the case, in particular,
for globular clusters and for the gaseous phase of fermionic matter
(at high energy and high temperature). The time required for a
metastable gaseous system to collapse is in general tremendously long
and increases exponentially with the number $N$ of particles (thus,
$t_{life}\rightarrow +\infty$ in the thermodynamic limit $N\rightarrow
+\infty$) \cite{meta}. This is due to the long-range nature of the
gravitational potential. Therefore, metastable states are in reality
stable states.  At high temperatures and high energies, the global
entropy maximum is not physically relevant
\cite{ko,ispolatov,grand,rieutord}. Condensed objects 
(e.g., planets, stars, white dwarfs, fermion balls,...)  only form
below a critical energy $E_{c}$ (Antonov energy)
\cite{antonov,lbw,paddy} or below a critical temperature $T_{c}$
(Jeans temperature) \cite{aa}, when the gaseous metastable phase
ceases to exist (spinodal point).

\subsection{The case $2<D<4$}
\label{sec_d3}

\begin{figure}[htbp]
\centerline{
\psfig{figure=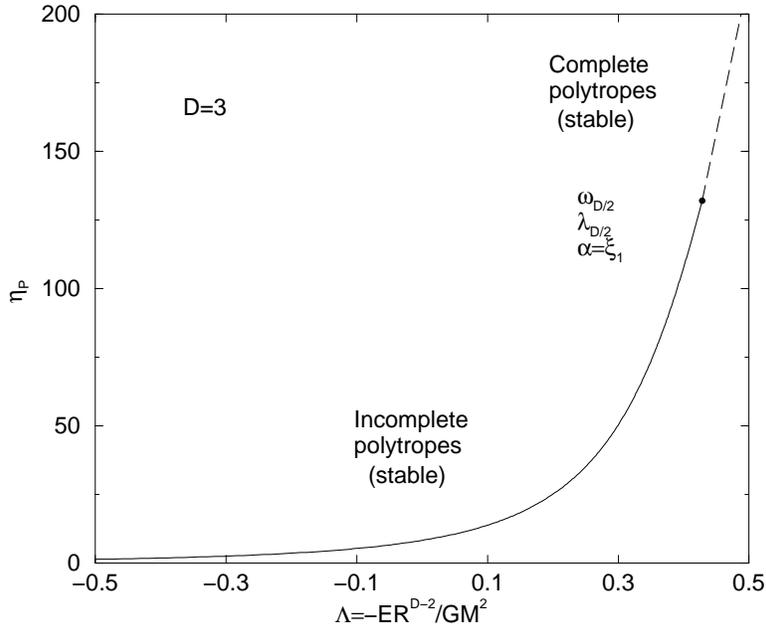,angle=0,height=8.5cm}}
\caption{The mass-energy relation for white dwarf stars ($T=0$) in $D=3$. There exists an equilibrium state for all mass. The white dwarf star is self-confined if $M>M_{*}(R)$ and box-confined if $M<M_{*}(R)$.}
\label{LHpD3}
\end{figure}

We start to describe the structure of the caloric curve of the
self-gravitating Fermi gas for $2<D<4$ (specifically $D=3$). Let us
first consider the Fermi gas at $T=0$ (white dwarf stars). The
$\Lambda-\eta_{P}$ curve defined by Eqs.  (\ref{d17}), (\ref{d18}) and
(\ref{d22}) is represented in Fig. \ref{LHpD3}. In the present
context, it gives the energy of the star as a function of its
mass. Since the curve does not present turning points, all the white
dwarf star configurations are stable. According to Eq. (\ref{d10}),
for $2<D<4$, the mass $M$ of a complete white dwarf star is a
decreasing function of its radius $R_*$. Therefore, if the system is
enclosed within a box, there exists a characteristic mass
\begin{equation}
M_*(R)={\chi_{D}\over \eta_{0}^{2\over D-2}G^{D\over
D-2}}R^{-{D(4-D)\over D-2}} \label{gtq1}
\end{equation}
such that for $M>M_*(R)$ the star is self-confined ($R_{*}<R$) and for
$M<M_*(R)$, it is restricted by the box. In terms of the dimensionless mass
$\eta_P$, complete $n_{3/2}$ polytropes correspond to $\eta_P\ge
\omega_{D/2}$ and incomplete $n_{3/2}$ polytropes to $\eta_P\le
\omega_{D/2}$. For $2<D<4$, there exists a stable
equilibrium at $T=0$ for all mass $M$.

\begin{figure}[htbp]
\centerline{
\psfig{figure=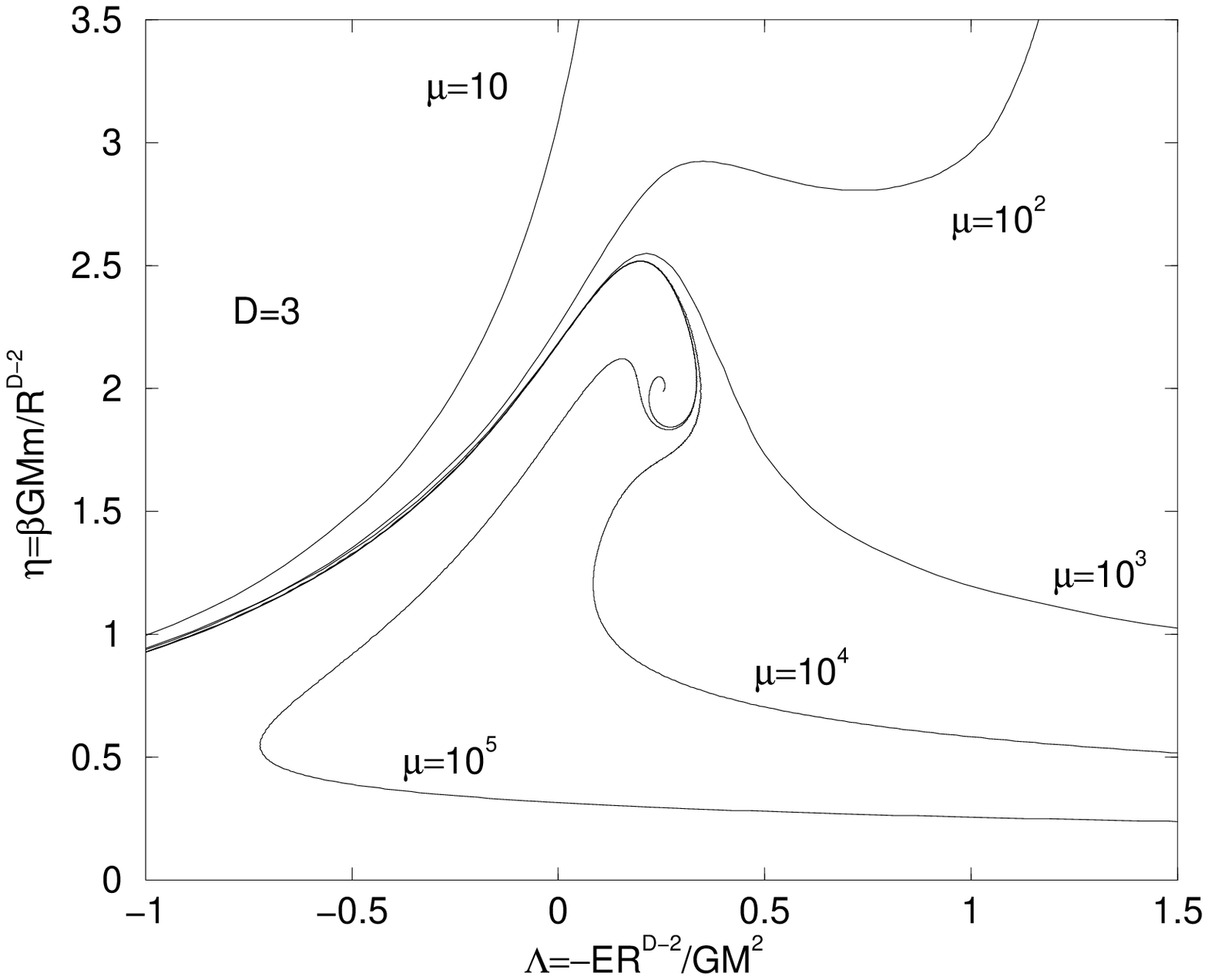,angle=0,height=8.5cm}}
\caption{Caloric curve in $D=3$ for different values of the degeneracy parameter (various system sizes).}
\label{LHD3}
\end{figure}

\begin{figure}[htbp]
\centerline{
\psfig{figure=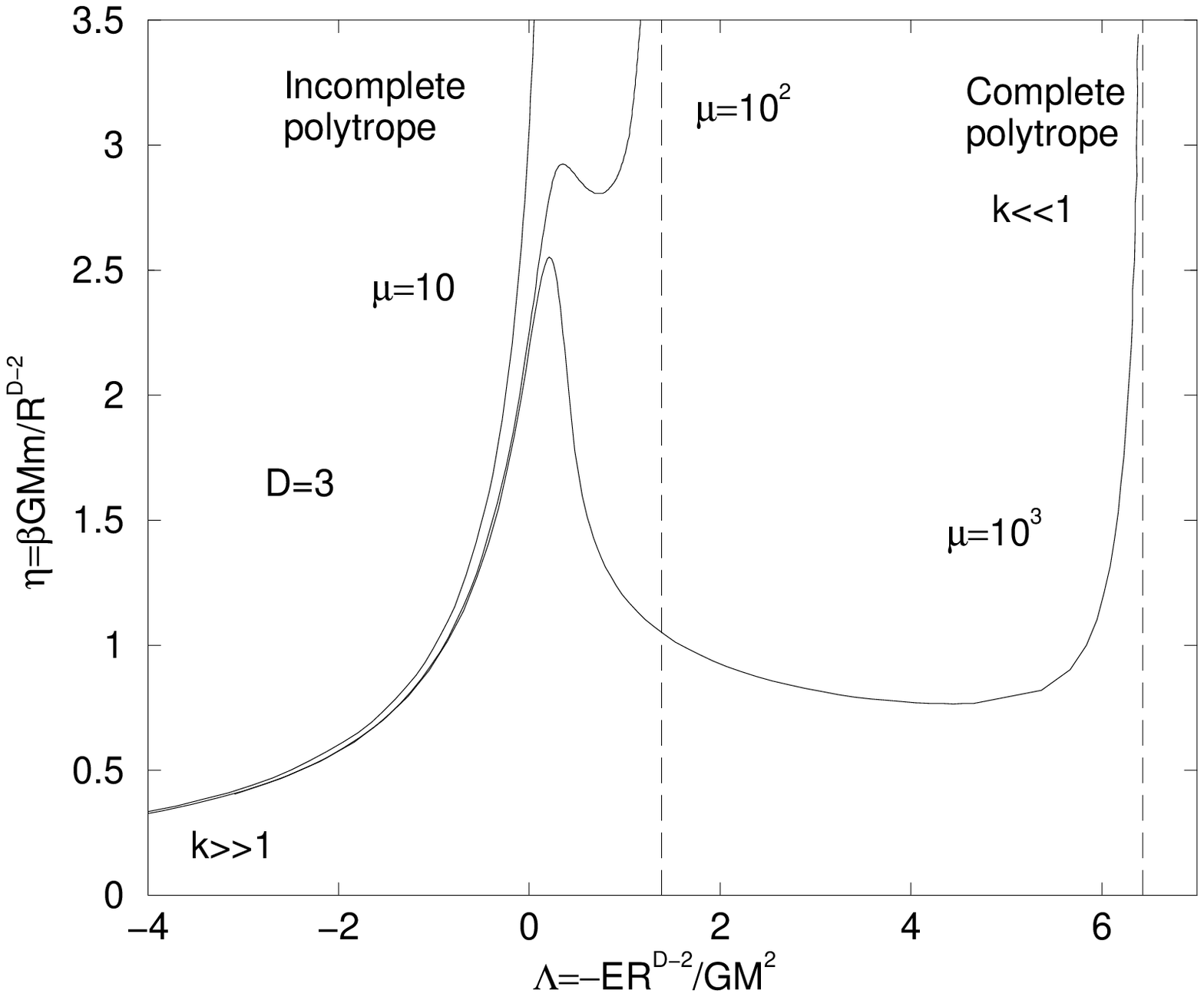,angle=0,height=8.5cm}}
\caption{Caloric curve in $D=3$ for small values of the degeneracy parameter (small system sizes). For $D<4$, there exists an equilibrium state for all temperatures $T$ and all accessible energies $E\ge E_{ground}$. }
\label{LHD3bis}
\end{figure}

We now briefly describe the caloric curve for arbitrary temperature
and energy. A more complete description is given in \cite{pt} for
$D=3$. First, we note that, according to Eqs. (\ref{p10}),
(\ref{d3}) and (\ref{d10}),
\begin{equation}
\mu=\mu_*(D) \biggl ({R\over R_*}\biggr )^{D(4-D)\over 2},
\label{gtq2}
\end{equation}
where
\begin{equation}
\mu_{*}(D)\equiv {D\over 2}(\omega_{D/2})^{D-2\over 2}.
\label{gtq3}
\end{equation}
Therefore, the degeneracy parameter $\mu$ can be seen as the ratio
(with some power) between the size of the system $R$ and the size
$R_{*}$ of a white dwarf star with mass $M$. Accordingly, a small
value of $\mu$ corresponds to a large ``effective'' cut-off (played by
Pauli's exclusion principle) or, equivalently, to a small system size.
Alternatively, a large value of $\mu$ corresponds to a small
``effective'' cut-off or a large system size. This gives a physical
interpretation to the degeneracy parameter. For $\mu\rightarrow
+\infty$ (i.e. $\hbar\rightarrow 0$), we recover classical isothermal
spheres. In that case, the caloric curve $\beta(E)$ forms a spiral.
For finite values of $\mu$, the spiral unwinds due to the influence of
degeneracy and gives rise to a rich variety of caloric curves
(Fig. \ref{LHD3}). For large systems, the caloric curve has a
$Z$-shape (``dinosaur's neck'') and for small systems it has a
$N$-shape. The phase transitions in the self-gravitating Fermi gas for
$D=3$ and the notion of metastable states, spinodal points, critical
points, collapse, explosion, and hysteresis are discussed in
\cite{pt,ispolatov,rieutord,meta}. Similar notions are discussed in \cite{stahl} for a hard spheres gas. The ground state of the self-gravitating
Fermi gas ($T=0$) corresponds to a white dwarf star configuration. For
given $\mu$, its structure (radius, energy) is determined by the
intersection between the $\Lambda-\eta_P$ curve in Fig. \ref{LHpD3}
and the line defined by Eq. (\ref{d19}). The ``white dwarf'' is
complete ($R_{*}<R$) for $\mu>\mu_{*}(D)$ and incomplete ($R_{*}>R$)
otherwise. For $\mu>\mu_{*}(D)$, the normalized energy of the white dwarf
is given by
\begin{equation}
\Lambda_{max}(D,\mu)=\lambda_{D/2}\biggl ({\mu\over\mu_{*}}\biggr )^{2(D-2)\over D(4-D)}.
\label{gtq4}
\end{equation}
This is the ground state of the self-gravitating Fermi gas
corresponding to the asymptote in Fig. \ref{LHD3bis} (this asymptote
exists for all curves in Fig. \ref{LHD3} but is outside the
frame). For classical particles ($\hbar=0$), there is no equilibrium
state if energy and temperature are below a critical threshold
\cite{antonov,lbw}. In that case, the system undergoes gravitational
collapse and forms binaries (in microcanonical ensemble) or a Dirac
peak (in canonical ensemble); see Appendices A and B of \cite{sc1} and
\cite{grand,rieutord,meta}. For self-gravitating fermions, an equilibrium
state exists for all values of temperature and for all accessible
energies ($E\ge E_{ground}$).  Gravitational collapse is arrested by
quantum pressure as first realized by Fowler
\cite{fowler}. We shall now show that this claim ceases to be true in
dimension $D\ge 4$.

\subsection{The case $4<D<2(1+\sqrt{2})$}
\label{sec_d41}

\begin{figure}[htbp]
\centerline{ \psfig{figure=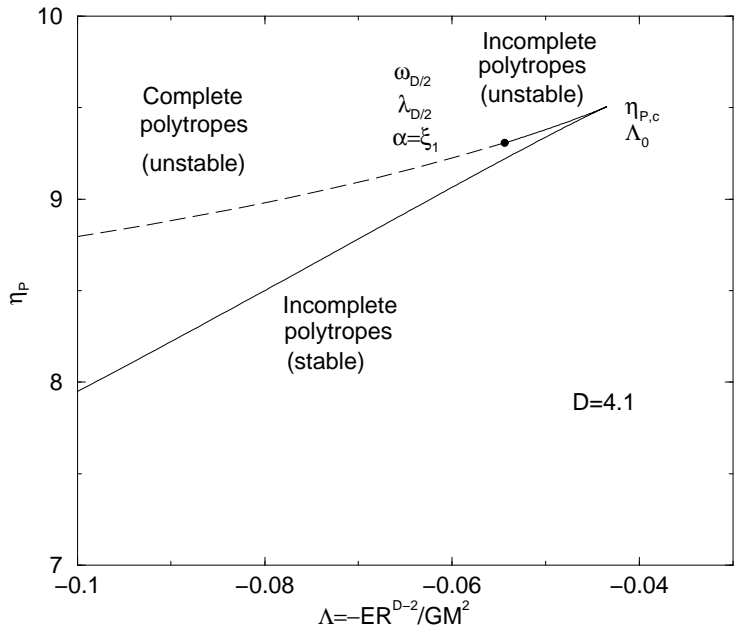,angle=0,height=8.5cm}}
\caption{The mass-energy relation for white dwarf stars ($T=0$) in $4<D<2(1+\sqrt{2})$ (specifically $D=4.1$). Self-confined white dwarf stars are always unstable. Box-confined white dwarf stars exist only for $M<M_{c}(R)$. For  $M>M_{c}(R)$, there is no equilibrium state.} \label{LHpD4.1}
\end{figure}

We now consider the case $4<D<2(1+\sqrt{2})$ (specifically
$D=4.1$). Let us first describe the Fermi gas at $T=0$. The
$\Lambda-\eta_{P}$ curve defined by Eqs.  (\ref{d17}), (\ref{d18}) and
(\ref{d22}) is represented in Fig. \ref{LHpD4.1}. For $D>4$, the
curves $\eta_{P}(\alpha)$ and $\Lambda(\alpha)$ associated to
$n_{3/2}$ polytropes have their extrema at the same point (see
Appendix C of \cite{langevin}). Therefore, the $\Lambda$-$\eta_{P}$
curve presents a cusp at $(\Lambda_{0},\eta_{P,c})$. Past this point
in the series of equilibria, $n_{3/2}$ polytropes are
unstable. According to Eq. (\ref{d10}), for $D>4$, the radius $R_*$ of
a self-confined white dwarf star increases with its mass. For
$M<M_*(R)$ there exists self-confined white dwarf star
configurations. In terms of the dimensionless mass $\eta_P$, this
corresponds to $\eta_P\le \omega_{D/2}$ (see
Fig. \ref{LHpD4.1}). However, such configurations are unstable since
they lie after the turning point \cite{langevin}. Therefore, only
incomplete (box confined) white dwarf stars can be stable in $D>4$. Inspecting
Fig. \ref{LHpD4.1} again, we observe that these configurations exist
only below a critical mass
\begin{equation}
M_c(R)=\eta_{P,c}(D) S_{D}R^{D(D-4)\over D-2}\biggl\lbrack
{K(D+2)\over 2S_{D}G}\biggr \rbrack^{D\over D-2}. \label{gtq5}
\end{equation}
For $M>M_c(R)$, there is no equilibrium state at $T=0$ for $D>4$. In terms
of the dimensionless mass $\eta_P$, equilibrium states exist only
for $\eta_{P}<\eta_{P,c}(D)$.

\begin{figure}[htbp]
\centerline{
\psfig{figure=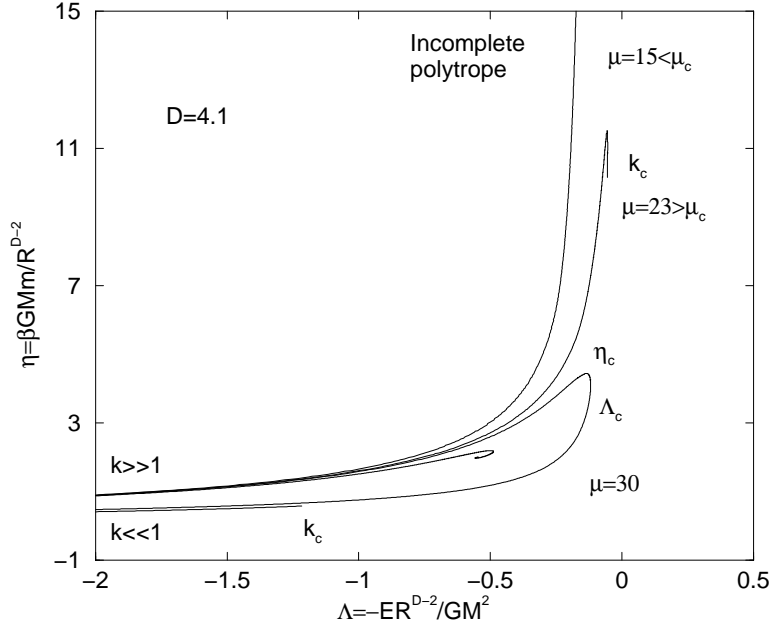,angle=0,height=8.5cm}}
\caption{Caloric curve in $D=4.1$ for different values of the degeneracy parameter. For $\mu>\mu_{c}(D)$, there is no equilibrium state if the  temperature and the energy are too low. The reason for the ``gap'' at $k_{c}$ is explained in Fig. \ref{test}.}
\label{LHD4.1}
\end{figure}

\begin{figure}[htbp]
\centerline{
\psfig{figure=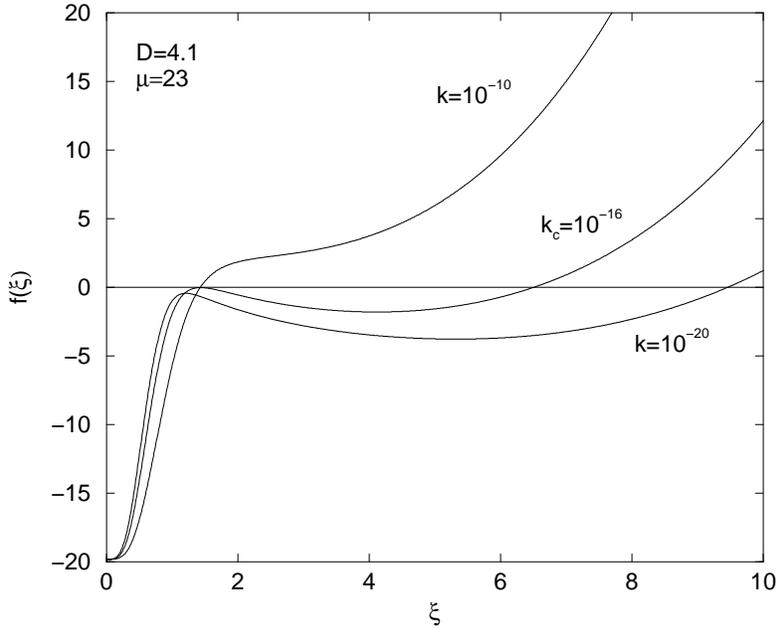,angle=0,height=8.5cm}}
\caption{Graphical construction determining the value of $\alpha$ for given $\mu$ and $k$ (in $D=4.1$). According to Eq. (\ref{p9}), the normalized box radius $\alpha$ is solution of $f(\alpha)=0$, where $f(\xi)=\xi^{D+2\over D-2}\psi'_{k}(\xi)-\mu^{2\over D-2}$. We see that $\alpha$ undergoes a discontinuity as $k\rightarrow k_{c}$. This gives rise to the ``gap'' in Fig. \ref{LHD4.1} for $\mu=23$. However, this gap is essentially a mathematical curiosity since the lower part of the curve (small $k$) is unstable anyway.  }
\label{test}
\end{figure}

The caloric curve for arbitrary value of temperature and energy is
represented in Fig. \ref{LHD4.1}. For $\mu\rightarrow +\infty$, we recover
the classical spiral \cite{sc1}. For finite values of $\mu$, there exists
equilibrium solutions at all temperatures only if
$\eta_P<\eta_{P,c}(D)$. Using Eq. (\ref{d19}), this corresponds to 
\begin{equation}
\mu<{D\over 2}\eta_{P,c}(D)^{D-2\over 2}\equiv \mu_{c}(D).
\label{gtq6}
\end{equation}
If $\mu>\mu_{c}(D)$, or equivalently if $M>M_*(R)$, there exists a
minimum energy $E_c=-\Lambda_c GM^2/R^{D-2}$ (which appears to be
positive) and a minimum temperature $T_c=GM/(\eta_c R^{D-2})$ below
which there is no equilibrium state (the values of $\eta_{c}$ and
$\Lambda_{c}$ depend on $D$ and $\mu$). In that case, the system is
expected to collapse. This is similar to the Antonov instability
(gravothermal catastrophe) for classical particles
\cite{antonov,lbw}. Since we deal here with self-gravitating fermions,
we could expect that quantum pressure would arrest the collapse. Our
study shows that this is not the case for $D>4$. Quantum mechanics
cannot stabilize matter against gravitational collapse anymore.

\subsection{The case $D=4$}
\label{sec_d4}

\begin{figure}[htbp]
\centerline{
\psfig{figure=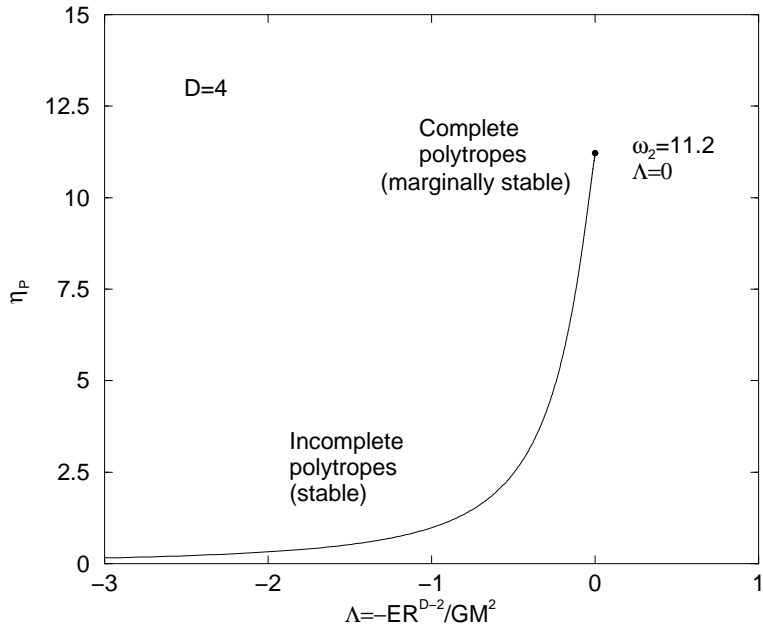,angle=0,height=8.5cm}}
\caption{The mass-energy relation for white dwarf stars ($T=0$) in $D=4$. Self-confined white dwarf stars are marginally stable. They have a unique mass $M_{limit}$ independent on their radius. For $M<M_{limit}$, the white dwarf star is box-confined. There is no equilibrium state with $M>M_{limit}$.}
\label{LHpD4}
\end{figure}

\begin{figure}[htbp]
\centerline{
\psfig{figure=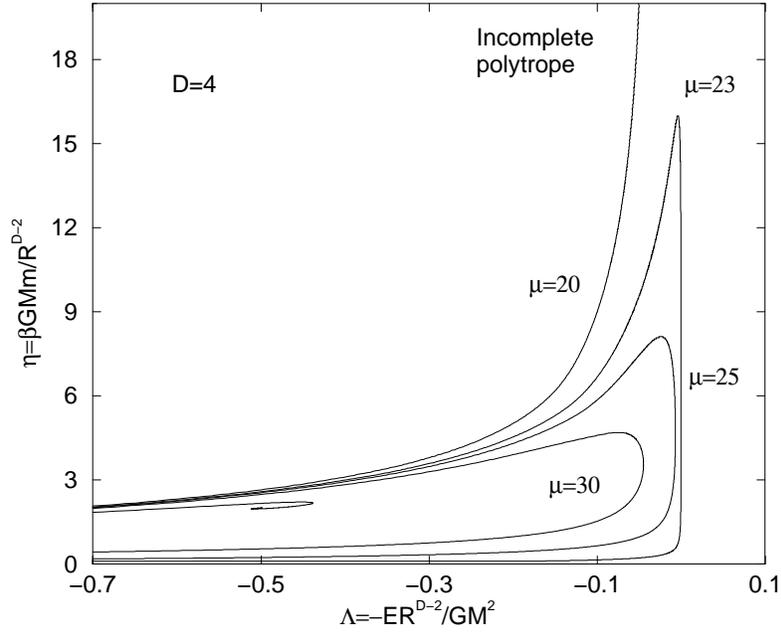,angle=0,height=8.5cm}}
\caption{Caloric curves in $D=4$ for different values of the degeneracy parameter. For $\mu>\mu_{c}=22.4$, there is no equilibrium state if the  temperature and the energy are too low. }
\label{LHD4}
\end{figure}

\begin{figure}[htbp]
\centerline{
\psfig{figure=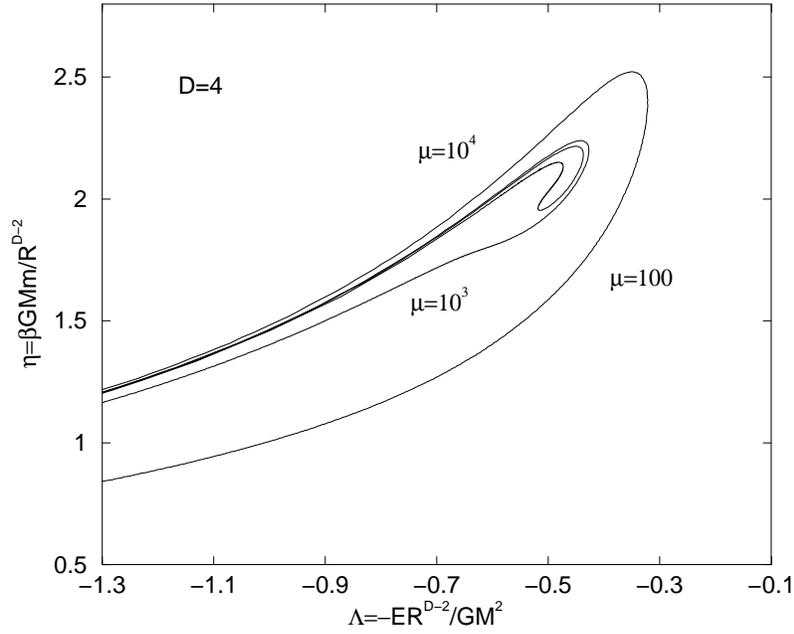,angle=0,height=8.5cm}}
\caption{Same as Fig. \ref{LHD4} for larger values of $\mu$ showing the developement of the classical spiral recovered for $\mu\rightarrow +\infty$.}
\label{LHD4ZOOM}
\end{figure}

The dimension $D=4$ is special because it is the dimension of space
above which quantum pressure cannot balance gravity anymore.
Therefore, $D=4$ is critical and it deserves a particular
attention. First, consider the Fermi gas at $T=0$. It corresponds to a
polytrope of index $n_{3/2}=n_3$ \cite{langevin}. The
$\Lambda-\eta_{P}$ curve defined by Eqs.  (\ref{d17}), (\ref{d18}) and
(\ref{d22}) is represented in Fig. \ref{LHpD4}. Since the curve is
monotonic the box-confined $n_{3/2}$ polytropes are stable and the
complete $n_{3/2}$ polytropes are marginally stable. For $D=4$, the
mass of a self-confined white dwarf star is independent on its radius,
see Eq. (\ref{d10}). It can be expressed in terms of fundamental
constants as
\begin{equation}
M_{limit}={\omega_{2}\over g S_{4}^{2}}{h^{4}\over
m^{5}G^{2}}\simeq 1.44\ 10^{-2}\ {h^{4}\over m^{5}G^{2}}, \label{gtq7}
\end{equation}
where $\omega_{2}\simeq 11.2$ (we have taken $g=2$ in the numerical
application). Mathematically, this is similar to Chandrasekhar's
limiting mass for relativistic white dwarf stars equivalent to $n=3$
polytropes in $D=3$ \cite{chandra1}. However, it is here a purely
classical (i.e. nonrelativistic) result. Relativistic effects will be
considered in a forthcoming paper \cite{relativity}. The energy of the
self-confined white dwarf stars is $E=0$. Considering Fig. \ref{LHpD4}
again, we see that incomplete white dwarf stars exist only for
$M<M_{limit}$. In terms of the dimensionless mass $\eta_P$, this
corresponds to $\eta<\eta_{P,c}=\omega_{2}\simeq 11.2$. For
$M>M_{limit}$, there is no equilibrium state at $T=0$. The caloric
curve for arbitrary value of temperature and energy is represented in
Fig. \ref{LHD4} (see an enlargement in Fig. \ref{LHD4ZOOM}). Its
description is similar to that of Sec \ref{sec_d41}. For
$M>M_{limit}$, or equivalently $\mu\ge \mu_{c}=2\omega_{2}\simeq
22.4.$ there exists a minimum energy $E_c=-\Lambda_c GM^2/R^{2}$ and a
minimum temperature $T_c=GM/(\eta_c R^{2})$ below which there is no
equilibrium state.

\subsection{The case $D\ge 2(1+\sqrt{2})$}
\label{sec_d48}

\begin{figure}[htbp]
\centerline{
\psfig{figure=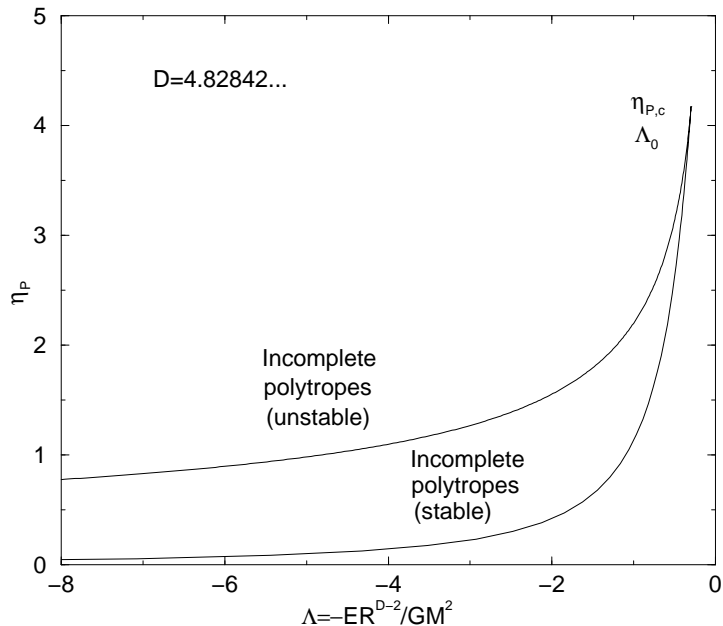,angle=0,height=8.5cm}}
\caption{The mass-energy relation for white dwarf stars ($T=0$) in $D=2(1+\sqrt{2})$. }
\label{LHpDn5}
\end{figure}

\begin{figure}[htbp]
\centerline{
\psfig{figure=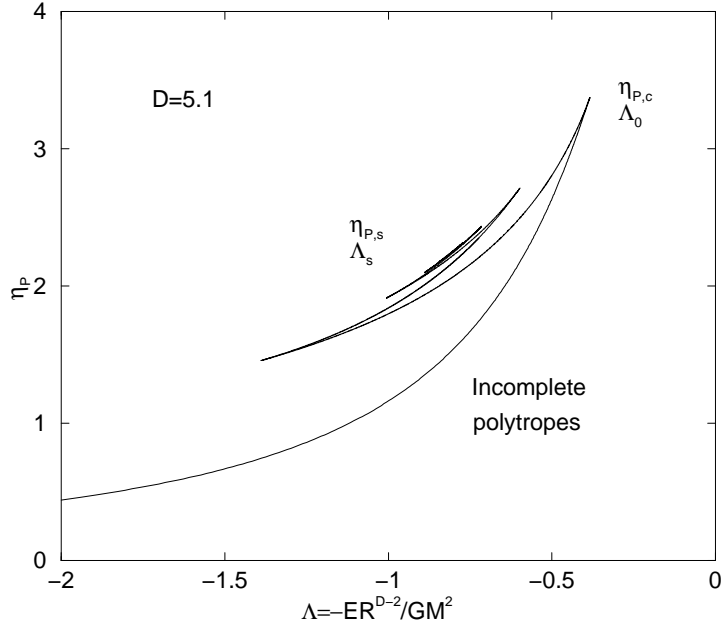,angle=0,height=8.5cm}}
\caption{The mass-energy relation for white dwarf stars ($T=0$) in $D=5.1$}
\label{LHpD5.1}
\end{figure}

The caloric curves for $D\ge 2(1+\sqrt{2})$ are similar to those of
Secs. \ref{sec_d41} and \ref{sec_d4}. There are, however, two main
differences. For $D\ge 10$, the classical spiral ceases to exist
\cite{sc1}. Thus, the caloric curve does not wind up as
$\mu\rightarrow +\infty$ contrary to Fig. \ref{LHD4ZOOM}.  On the
other hand, for $D\ge 2(1+\sqrt{2})$, it is not possible to construct
self-confined white dwarf stars
\cite{langevin}.  This is just a mathematical curiosity since complete
white dwarfs stars are unstable for $D>4$ anyway. This property
changes the unstable branch of the $\Lambda-\eta_{P}$ diagram without
consequence on the caloric curves. The $\Lambda-\eta_{P}$ diagram is
represented Figs. \ref{LHpDn5} and
\ref{LHpD5.1}. For $D>2(1+\sqrt{2})$, it displays an infinity of
cusps towards the singular solution ($\Lambda_{s}$,$\eta_{P,s}$), see
Fig. \ref{LHpD5.1}. For $D=2(1+\sqrt{2})$, there is just one cusp (see
Fig. \ref{LHpDn5}) and the Lane-Emden equation (\ref{d7}) can be solved
analytically.  This corresponds to the $D$-dimensional Schuster
solution obtained for $n=n_{5}$
\cite{langevin}. In that case, we find explicitly
\begin{equation}
\theta_{5}={1\over \bigl \lbrack 1+{\xi^{2}\over 4(2+\sqrt{2})}\bigr \rbrack^{\sqrt{2}}}.
\label{fw1}
\end{equation}
The normalized mass and the normalized energy can be expressed as
\begin{equation}
\eta_{P}={\alpha^{2+\sqrt{2}}\over 2(1+\sqrt{2})\bigl \lbrack 1+{\alpha^{2}\over 4(2+\sqrt{2})}\bigr \rbrack^{1+\sqrt{2}}},
\label{fw2}
\end{equation}
\begin{equation}
\Lambda_{5}=-2(1+\sqrt{2})\biggl\lbrack 1+{\alpha^{2}\over 4(2+\sqrt{2})}\biggr\rbrack^{2(1+\sqrt{2})}{1\over \alpha^{2(2+\sqrt{2})}}\int_{0}^{\alpha}{\xi^{1+2\sqrt{2}}\over \bigl\lbrack 1+{\xi^{2}\over 4(2+\sqrt{2})}\bigr\rbrack^{2(1+\sqrt{2})}}d\xi.
\label{fw3}
\end{equation}

\subsection{The case $D=2$}
\label{sec_d2}

Let us now consider smaller dimensions of space. The dimension $D=2$
is critical concerning gravitational collapse as discussed in \cite{sc1}. For
$D=2$, the relevant Fermi integrals are $I_{0}$ and $I_{1}$. By
definition,
\begin{equation}
I_{0}(t)=\int_{0}^{+\infty}{dx\over 1+t e^{x}}. \label{t1}
\end{equation}
Changing variables to $y=e^{x}$, we easily find that
\begin{equation}
I_{0}(t)=\ln\biggl (1+{1\over t}\biggr ).
\label{t2}
\end{equation}
Therefore, the Fermi-Poisson equation (\ref{p4}) becomes
\begin{equation}
{1\over\xi}{d\over d\xi}\biggl (\xi {d\psi\over d\xi}\biggr )=\ln\bigl (1+k^{-1}e^{-\psi}\bigr ).
\label{t3}
\end{equation}
\begin{equation}
\psi(0)=\psi'(0)=0.
\label{t4}
\end{equation}
On the other hand, using the identity (\ref{p3}), giving
\begin{equation}
I'_{1}(t)=-{1\over t}\ln\biggl (1+{1\over t}\biggr ),
\label{t5}
\end{equation}
one finds that
\begin{equation}
I_{1}(t)=-\int_{-{1/t}}^{0}{\ln(1-x)\over x}dx=-{\rm Li}_{2}\bigl (-{1\over t}\bigr ),
\label{t6}
\end{equation}
where ${\rm Li}_{2}$ is the dilogarithm.

Consider first the Fermi gas at $T=0$. In $D=2$, a white dwarf star is
equivalent to a polytrope with index $n_{3/2}=1$. The Lane-Emden
equation can then be solved analytically and we obtain
$\theta=J_{0}(\xi)$, where $J_{0}$ is the Bessel function of zeroth
order.  The density drops to zero at $\xi_{1}=\alpha_{0,1}\simeq
2.40$, the first zero of $J_{0}$.  Considering the mass-radius
relation (\ref{d10}) in $D=2$, we see that the radius is independant
on mass. Therefore, complete white dwarf stars in two dimensions all
have the same radius. It can be written in terms of fundamental
constants as
\begin{equation}
R_{*}={\xi_1\over 2\pi}\biggl ({h^{2}\over g m^{3}G}\biggr
)^{1/2}=0.27\ {h\over m^{3/2}G^{1/2}}. \label{t7}
\end{equation}
The relation between the mass and the central density of the white
dwarf star is
\begin{equation}
M={\rho_{0}\over 4\pi^{2}}{h^{2}\over g m^{3}G}\xi_{1}|\theta'_{1}|,
\label{t8}
\end{equation}
where $\theta'_{1}=J_{0}'(\alpha_{0,1})\simeq -0.52$. Thus, the
density profile of a two-dimensional white dwarf star can be written
\begin{equation}
\rho(r)=\rho_{0}J_{0}\bigl ({\xi_{1}r\over R_{*}}\bigr ).
\label{t9}
\end{equation}
This is similar to the vorticity profile of a minimum enstrophy
vortex in 2D hydrodynamics \cite{jfm,gt}. The energy of a complete polytrope of
index $n$ in $D=2$ is $E=-(n-1)GM^{2}/8+(1/2)GM^{2}\ln (R_{*}/R)$
with the convention $\Phi(R)=0$ \cite{langevin}. Therefore the energy of a 2D
white dwarf star  is
\begin{equation}
E={1\over 2}GM^{2}\ln \bigl ({R_{*}\over R}\bigr ).
\label{t10}
\end{equation}
Two-dimensional white dwarf stars exist for any mass $M$ and they
are stable. Noting that
$R_{*}/R=(\mu_{*}/\mu)^{1/2}=\xi_1/\sqrt{\mu}$ where
$\mu=4\pi^{2}\eta_{0}GR^{2}$, we can write the normalized energy
of the self-confined white dwarf star as
\begin{equation}
\Lambda={1\over 2}\ln\biggl ({\sqrt{\mu}\over\xi_1}\biggr ).
\label{t11}
\end{equation}
Let us now consider the case of incomplete white dwarf stars that are
confined by the box ($R_*>R$). This corresponds to
$\mu<\xi_1^2$. Using Eq. (\ref{d6}), we find that
$\alpha=\sqrt{\mu}$. Then, using the results of
\cite{langevin}, we find that the normalized energy of a box-confined
white dwarf star in two dimensions is
\begin{equation}
\Lambda=-{1\over
2}{J_{0}(\sqrt{\mu})\over\sqrt{\mu}J_{1}(\sqrt{\mu})}. \label{t12}
\end{equation}

\begin{figure}[htbp]
\centerline{
\psfig{figure=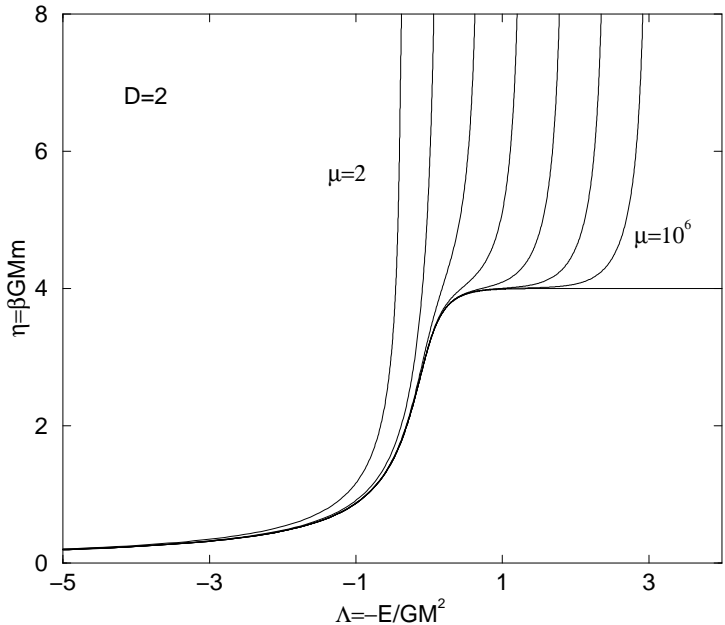,angle=0,height=8.5cm}}
\caption{Caloric curve in $D=2$ for different values of the degeneracy parameter: $\mu=2,10,100,10^3,10^4,10^5,10^6$. For $\mu\rightarrow +\infty$, we recover the classical caloric curve displaying a critical temperature $T_{c}$. Below $T_{c}$ the system is expected to collapse and create a Dirac peak (``black hole''). When quantum mechanics is accounted for, the ``black hole'' is replaced by a ``fermion ball''. This result is generally valid for $2\le D<4$.}
\label{LHD2}
\end{figure}

We now consider the self-gravitating Fermi gas at finite
temperature $T\neq 0$. According to Eq. (\ref{p6}) we have
$\alpha=\sqrt{\mu}$. Using Eq. (\ref{p8}), we obtain
\begin{equation}
\eta\equiv \beta GMm=\sqrt{\mu}\ \psi'(\sqrt{\mu}).
\label{t12bis}
\end{equation}
We need to calculate the energy specifically because the expression
(\ref{p16}) breaks down in $D=2$. The kinetic energy $K=\int p
d^{2}{\bf r}$ can be written
\begin{equation}
{K\over GM^{2}}={1\over \eta^{2}}\int_{0}^{\sqrt{\mu}}I_{1}(k e^{\psi})\xi d\xi.
\label{t13}
\end{equation}
On the other hand, using an integration by parts, the potential energy is
given by
\begin{equation}
W=-{1\over 4\pi G}\int (\nabla\Phi)^{2}d^{2}{\bf r}, \label{t14}
\end{equation}
where we have taken $\Phi(R)=0$. Introducing the
dimensionless quantities defined in Sec. \ref{sec_para}, we get
\begin{equation}
{W\over GM^{2}}=-{1\over 2\eta^{2}}\int_{0}^{\sqrt{\mu}}\psi'(\xi)^{2}\xi d\xi.
\label{t15}
\end{equation}
Summing Eqs. (\ref{t13}) and (\ref{t15}), the total normalized energy of the Fermi gas
in two dimensions is
\begin{equation}
\Lambda\equiv -{E\over GM^{2}}=-{1\over \eta^{2}}\int_{0}^{\sqrt{\mu}}I_{1}(k e^{\psi})\xi d\xi+{1\over 2\eta^{2}}\int_{0}^{\sqrt{\mu}}\psi'(\xi)^{2}\xi d\xi.
\label{t16}
\end{equation}
The corresponding caloric curve is plotted in Fig. \ref{LHD2}. For
$\mu\rightarrow +\infty$, we recover the classical caloric curve
displaying a critical temperature $k_{B}T_{c}=GMm/4$ \cite{sc1}. Below
$T_{c}$, a classical gas experiences a gravitational collapse and
develops a Dirac peak \cite{sc1}. When quantum mechanics is taken into
account, the collapse stops when the system becomes degenerate. The
Dirac peak is replaced by a fermion ball surrounded by a dilute
halo. At $T=0$, we have a pure Fermi condensate without halo. This is
the ground state of the self-gravitating Fermi gas corresponding to
the vertical asymptotes in Fig.  \ref{LHD2}. For $\mu<\xi_1^2$
(incomplete white dwarf stars), the minimum energy is given by
Eq. (\ref{t12}) and for $\mu<\xi_1^2$ (complete white dwarf stars) by
Eq. (\ref{t11}). This discussion concerning the difference between
Dirac peaks (for classical particles) and fermion balls (for quantum
particles) in the canonical ensemble remains valid for
$2\le D<4$. Note also that there is no collapse (gravothermal
catastrophe) in the microcanonical ensemble in $D=2$ \cite{klb,sc1}.

\subsection{The case $D<2$}
\label{sec_d1}

\begin{figure}[htbp]
\centerline{
\psfig{figure=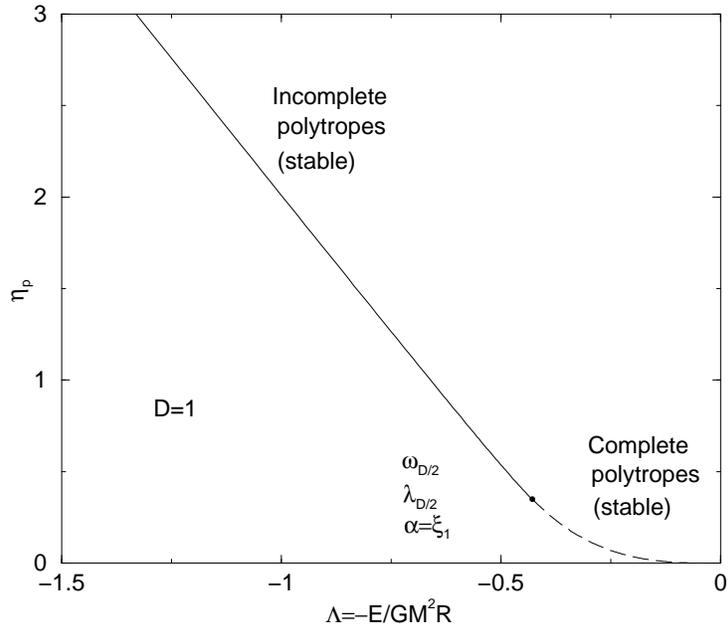,angle=0,height=8.5cm}}
\caption{The mass-energy relation for white dwarf stars ($T=0$) in $D<2$ (specifically $D=1$). There exists an equilibrium state for all mass. The white dwarf star is self-confined if $M<M_{*}(R)$ and box-confined if $M>M_{*}(R)$.}
\label{LHpD1}
\end{figure}

We finally conclude by the case $D<2$ (specifically $D=1$).  First, we
consider the Fermi gas at $T=0$. The $\Lambda-\eta_{P}$ curve which
gives the energy of the star as a function of its mass is represented
in Fig. \ref{LHpD1}. Since the curve does not present turning points,
all the white dwarf star configurations are stable. According to
Eq. (\ref{d10}), for $D<2$, the mass $M$ of a complete white dwarf star
increases with its radius $R_*$. Therefore, for $M<M_*(R)$ the star is
self-confined and for $M>M_*(R)$ it is restricted by the box.  There
exists a stable equilibrium state at $T=0$ for all mass. In terms of the
dimensionless mass $\eta_P$, complete $n_{3/2}$ polytropes correspond
to $\eta_P\le \omega_{D/2}$ and incomplete $n_{3/2}$ polytropes to
$\eta_P\ge \omega_{D/2}$. This situation is reversed with respect to
that of Fig. \ref{LHpD3}.

\begin{figure}[htbp]
\centerline{
\psfig{figure=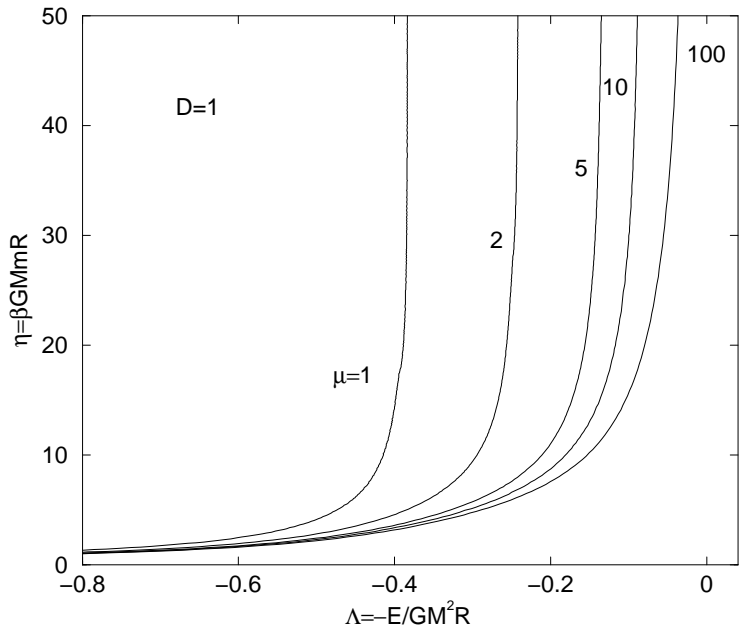,angle=0,height=8.5cm}}
\caption{Caloric curve in $D=1$ for different values of the degeneracy parameter (various system sizes). }
\label{LHD1}
\end{figure}

The caloric curve for arbitrary temperature and energy is represented
in Fig. \ref{LHD1}. For $\mu\rightarrow +\infty$ (i.e.
$\hbar\rightarrow 0$), we recover the curve obtained in \cite{sc1} for
classical isothermal systems.  The caloric curve $\beta(E)$ is
monotonic.  Therefore, there is no phase transition for $D<2$. Thus,
the change in the caloric curve due to quantum mechanics is not very
important since an equilibrium state (global maximum of entropy or
free energy) already exists for any accessible energy $E$ and any
temperature $T$ in classical mechanics. Quantum mechanics, however,
changes the ground state of the system. The ground state of the
self-gravitating Fermi gas ($T=0$) corresponds to a white dwarf star
configuration. Its structure (radius, energy) is determined by the
intersection between the $\Lambda-\eta_P$ curve in Fig. \ref{LHpD1}
and the line defined by Eq. (\ref{d19}). The ``white dwarf'' is
complete ($R_{*}<R$) for $\mu>\mu_{*}(D)$ and incomplete ($R_{*}>R$)
otherwise. For $\mu>\mu_{*}(D)$, the normalized energy of the white
dwarf is given by Eq. (\ref{gtq4}). This is the ground state of the
self-gravitating Fermi gas corresponding to the asymptote in
Fig. \ref{LHD1}. In $D=1$, it is possible to obtain more explicit
results. Using the results of
\cite{langevin}, for $n_{3/2}=1/2$ polytropes, we have
$\xi_{1}=(3\pi/4)^{1/2}\Gamma(5/3)/\Gamma(7/6)\simeq 1.49$ and
$|\theta'_{1}|=2/\sqrt{3}\simeq 1.15$. Therefore,
$\omega_{1/2}=0.349$ and $\mu_{*}=0.846$. For $\mu>\mu_{*}=0.846$,
the normalized energy of a complete white dwarf star (ground state) is
\begin{equation}
\Lambda_{min}=-{3\over 7}\biggl ({\mu_{*}\over \mu}\biggr )^{2/3}.
\label{bff}
\end{equation}

\section{Conclusion} \label{sec_conclusion}

In this paper, we have studied how the dimension of space affects the
nature of phase transitions in the self-gravitating Fermi gas. Since
this model has a fundamental interest in astrophysics \cite{chavcape}
and statistical mechanics \cite{houches}, it is important to explore
its properties thoroughly even if we sacrifice for practical
applications. It is well-known in statistical mechanics that the
dimension of space plays a crucial role in the problem of phase
transitions. For example, concerning the Ising model, the behaviour in
$D=1$ and $D\ge 2$ is radically different
\cite{huang}. We have reached a similar conclusion for the
self-gravitating Fermi gas. The solution of the problem in $D<2$ does
not yield any phase transition. In $D=2$, phase transitions appear in
the canonical ensemble but not in the microcanonical ensemble. In
$D>2$, phase transitions appear both in microcanonical and canonical
ensembles in association with gravitational collapse. The beauty of
self-gravitating systems, and other systems with long-range
interactions, is their simplicity since the mean-field approximation
is exact in any dimension. Therefore, the mean-field theory does {\it
not} predict any phase transition for the self-gravitating Fermi gas
in $D=1$, contrary to the Ising model.

At a more philosophical level, several scientists have examined the
role played by the dimension of space in determining the form of the
laws of physics. This question goes back to Ptolemy who argues in his
treatise {\it On dimensionality} that no more than three spatial
dimensions are possible in Nature. In the $18^{\rm th}$ century, Kant
realizes the deep connection between the inverse square law of
gravitation and the existence of three spatial dimensions. In the
twentieth century, Ehrenfest \cite{ehrenfest} argues that planetary
orbits, atoms and molecules would be unstable in a space of dimension
$D\ge 4$. Other investigations on dimensionality are reviewed in the
paper of Barrow \cite{barrow}.  Although we ignored this literature at
the begining, our study clearly enters in this type of
investigations. We have found that the self-gravitating Fermi gas
possesses a rich structure and displays several characteristic
dimensions $D=2$, $D=4$, $D=2(1+\sqrt{2})$ and $D=10$. Moreover, as
already noted in \cite{langevin}, the dimension $D=4$ is critical
because at that dimension quantum mechanics cannot stabilize matter
against gravitational collapse, contrary to the situation in
$D=3$. Interestingly, this result is similar to that of Ehrenfest
although it applies to white dwarf stars instead of atoms. The
dimension $D=2$ is also critical as found in \cite{sc1} and in
different domains of physics. Therefore, the dimension of our
(macroscopic) universe $D=3$ plays a very special role regarding the
laws of physics (this is illustrated in Fig. \ref{MRD}). Following the
far reaching intuition of Kant, we can wonder whether the three space
dimensions are a consequence of Newton's inverse square law, rather
than the opposite. We note also that extra-dimensions can appear at
the microscale, an idea originating from Kaluza-Klein theory. This
idea took a renaissance in modern theories of grand unification. Our
approach shows that already at a simple level, the coupling between
Newton's equations (gravitation) and Fermi-Dirac statistics (quantum
mechanics) reveals a rich structure as a function of $D$. Relativistic
effects will be considered in a forthcoming paper \cite{relativity}.

Finally, our study can shed light on the mathematical properties of
the Vlasov-Poisson system. Indeed, there is a close connexion between
collisionless stellar systems and self-gravitating fermions
\cite{lb,csr,cs,dubrovnik}. For example, the fact that the Vlasov equation
does not blow up (i.e., experiences gravitational collapse) in $D=3$
for non singular initial conditions can be related to a sort of exclusion
principle, as in quantum mechanics. Due to the Liouville theorem in
$\mu$-space, the distribution function must remain smaller that its
maximum initial value $f\le \eta_{0}$ and this prevents complete
collapse \cite{cs,robert}, unlike for collisional stellar systems
\cite{lbw} described by the Landau-Poisson system. Since quantum
mechanics cannot arrest gravitational collapse in $D\ge 4$ (for
sufficiently low energies), this suggests that the Vlasov-Poisson
system can probably blow up for $D\ge 4$. This remark could be of
interest for mathematicians.

\vskip2cm

{\bf Acknowledgements} I acknowledge interesting discussions with
P. Biler, T. Nadzieja and R. Sta\'nczy. I also thank B. Dou\c cot for
encouragements.

\end{document}